\documentstyle[12pt,cjp]{article}

\newcommand{\be}{\begin{equation}}
\newcommand{\ee}{\end{equation}}
\newcommand{\ba}{\begin{eqnarray}}
\newcommand{\ea}{\end{eqnarray}}

\newcommand{\ga}{\gamma_5}
\newcommand{\dg}{^{\dagger}}
\newcommand{\re}[1]{(\ref{#1})}

\newcommand{\Id}{\mbox{1\hspace{-0.98mm}l}}   
\newcommand{\mb}[1]{\quad\mbox{ #1 }\quad}
\newcommand{\e}{\mbox{e}}
\newcommand{\di}{\mbox{d}\,}

\newcommand{\la}{\lambda}
\newcommand{\cao}{{\cal O}} 
 
\newcommand{\pt}{\partial} 
 
\newcommand{\G}{\Gamma} 
\newcommand{\bG}{\bar{\Gamma}} 
\newcommand{\df}{[\mbox{d}\bar{\psi}\mbox{d}\psi]}
\newcommand{\dfp}{[\mbox{d}\bar{\psi}'\mbox{d}\psi']}
\newcommand{\f}{{\mbox{\tiny f}}} 

\begin{document}

\vspace*{-8mm}
\begin{flushright} {\sc HUB-EP}-99/66 \end{flushright} 
\vspace*{5mm}

\title{Dirac operator normality and chiral fermions
\footnote{Talk at CHIRAL '99, Taipei, Taiwan, Sept.~13-18, 1999.}
}
\author{Werner Kerler}
\address{Institut f\"ur Physik, Humboldt-Universit\"at, D-10115 Berlin, 
Germany}

\maketitle

\begin{abstract}
Normality of the Dirac operator is shown to be necessary for chiral properties.
 From the global chiral Ward identity, which in the continuum limit gives the 
index theorem, a sum rule results which constrains the spectrum. The 
Ginsparg-Wilson relation is to be restricted to its simple form and is a member
of a set of spectral constraints. A family of alternative chiral 
transformations is introduced. The one of L\"uscher is a special case which 
transports only the anomaly term to the measure. An alternative transformation 
would also be needed to correct Fujikawa's path-integral approach. From a 
general function of the hermitean Wilson-Dirac operator the one of Neuberger 
follows.
\end{abstract}

\section{Introduction}

Imposing the Ginsparg-Wilson (GW) relation \cite{gi82} in 
Refs.~\cite{ha98,lu98} a particular formula for the index of the massless
Dirac operator $D$ on the lattice has been given. Chiu \cite{ch98} has observed
that with the simple form of the GW relation and $\ga$-hermiticity the Dirac 
operator $D$ gets normal and its index and the corresponding difference at its 
second real eigenvalue add up to zero. This raises the questions what 
precisely the general conditions for chiral properties are and which general 
rules follow.

Assuming the GW relation, L\"uscher \cite{lu98} has introduced an alternative 
chiral transformation under which the measure is no longer invariant, a 
generalized finite form of which has been given by Chiu \cite{ch99}. This 
reminds of an old claim by Fujikawa \cite{fu79} in continuum theory 
that the chiral anomaly could be obtained from the measure. 
Thus it appears appropriate now to clarify in general what the r\^ole of the 
integration measure is. 

Neuberger \cite{ne98} has derived an explicit form of the massless Dirac 
operator on the lattice from the overlap formalism \cite{na93}. It is of 
interest whether this form, which relies on the hermitean Wilson-Dirac operator
$H$, also follows from other general requirements and under which conditions it 
is the only solution of the GW relation. 

In the following we start from the general Ward identity holding in a 
background gauge field. In this context we also introduce a family of 
alternative chiral transformations which give the same Ward identity, 
however, allow to transport terms between the action contribution and 
the measure contribution.

Next we show that it is necessary, in addition to $\ga$-hermiticity, to have 
normality of the Dirac operator $D$ in order to get chiral eigenstates and 
thus chiral properties. We derive general consequences for the terms in the 
Ward identity, which then gets the general sum rule for chiral differences. 
This rule is seen to put severe restrictions on the spectrum of $D$ which are 
crucial for allowing a nonvanishing index.

With respect to the GW relation we notice that its general form does not 
guarantee normality of $D$ so that we have to restrict to its simple form. We 
observe that, given $\ga$-hermiticity, this relation is actually a spectral 
constraint. Using a decomposition of $D$ we give an example of a family of 
spectral constraints of which the GW one is a member. The alternative 
transformation which transports the anomaly term to the measure in the GW case 
is seen to get that of L\"uscher. 

We point out that in the continuum limit the index theorem follows
from the lattice Ward identity and one has still Tr$(\ga)=0$. With different 
origins of the anomaly term, there is agreement with the expectations of 
conventional continuum theory. However, the derivation of Fujikawa's 
path-integral approach turns out not to agree with what is well defined from 
the lattice. A correction of this approach would include the use of an 
alternative transformation.

To study what follows starting from the hermitean Wilson-Dirac operator $H$ 
we require $D$ to be normal, $\ga$-hermitean, and a general function of $H$. 
This is seen to lead to the operator of Neuberger. It also establishes the 
connection to the GW relation.

\section{Ward identities} \label{Wd}

Fermionic Ward identities arise from  the condition that 
$\int \df \e^{-S_\f} \cao$ must not change under a transformation of the 
integration variables. Considering in particular the transformation 
$\psi'= \exp(i\eta\G)\psi$, $\bar{\psi}'= \bar{\psi} \exp(i\eta\bar{\G})$
this can be expressed by the identity 
\be
\frac{\di}{\di\eta} \int\dfp \e^{-S_\f'} \cao' \Big|_{\eta=0} = 0 
\label{w0s}
\ee
where $S_\f'=\bar{\psi'}M\psi'$. Evaluation of \re{w0s} gives 
\be
i\int\df \e^{-S_\f}\Big(-\mbox{Tr}(\bar{\G}+\G)\cao 
-\bar{\psi}(\bar{\G}M+M\G)\psi\cao + 
\bar{\psi}\bG\frac{\pt \cao}{\pt \bar{\psi}} -
\frac{\pt \cao}{\pt \psi}\G\psi \Big) = 0 \; ,
\label{w1s}
\ee 
with three contributions, one from the derivative of the 
integration measure, one from that of the action, and one from that of $\cao$.

In the present context one usually puts $\cao=1$. We can, however, do better
integrating out the $\psi$ and $\bar{\psi}$ fields in the second term of 
\re{w1s}, which after a calculation relying on Grassmann properties gives 
\be
 i\;\mbox{Tr}\Big(-\bar{\G}-\G + M^{-1}(\bar{\G}M+M\G)\Big) 
 \int\df \e^{-S_\f}\cao =0 \;.
\label{iW}
\ee
 From \re{iW} it 
becomes obvious that in a background gauge field the expectation value 
factorizes so that also for arbitrary $\cao$ (and not only for $\cao=1$) it 
suffices to consider the identity 
\be
 \frac{1}{2}\mbox{Tr}\Big(-\bar{\G}-\G + M^{-1}(\bar{\G}M+M\G)\Big)=0 
\label{W0}
\ee
where $-\frac{1}{2}\mbox{Tr}(\bG+\G)$ is the measure contribution and 
$\frac{1}{2}\mbox{Tr}\Big(M^{-1}(\bar{\G}M+M\G)\Big)$ the action contribution.  

For the global chiral transformation, in which case one has $\G=\bar{\G}=\ga$,
the measure contribution vanishes and \re{W0} becomes 
\be
\frac{1}{2}\mbox{Tr}(M^{-1}\{\ga,M\})=0 \;.
\label{WM}
\ee 
Obviously this can also be read as Tr$(\ga)=0$, of which the Ward identity is 
the particular decomposition which is dictated by the chiral transformation. 

In the presence of zero modes of a Dirac operator $D$ it is crucial to take
care that the derived relations remain properly defined. To guarantee this
we put $M=D+\varepsilon$ and let $\varepsilon$ go to zero in the final result. 
Thus from \re{WM} we altogether have
\be
\mbox{Tr}(\ga) = \frac{1}{2}\mbox{Tr}\Big((D+\varepsilon)^{-1}\{\ga,D\}\Big)+
\varepsilon\mbox{Tr}\Big((D+\varepsilon)^{-1}\ga\Big)=0 
\label{gwa2}
\ee
with $\ga$, of course, understood as $\ga$ times the appropriate unit operator. 
To have definite names in our discussions we shall call the first term in 
\re{gwa2}, which only contributes if $\{\ga,D\}\ne 0$, anomaly term, and the 
second one mass term or index term (for $\varepsilon\rightarrow0$). The 
operator $D$ is considered to be massless. 

We note that a family of alternative global chiral transformations can be
introduced putting $\G=\ga-K$, $\bar{\G}=\ga-\bar{K}$, which inserted 
into \re{W0} gives 
\be
-\frac{1}{2}\mbox{Tr}(\bG+\G)=+\frac{1}{2}\mbox{Tr}(K+\bar{K})
\label{mes}
\ee
for the measure contribution and 
\be
\frac{1}{2}\mbox{Tr}\Big(M^{-1}(\bG M+M\G)\Big)=
\frac{1}{2}\mbox{Tr}(M^{-1}\{\ga,M\})-\frac{1}{2}\mbox{Tr}(K+\bar{K})
\label{act}
\ee
for the action contribution. Obviously the extra term of the latter cancels 
the measure term so that again the result \re{WM} is obtained for any 
operators $K$ and $\bar{K}$.

While the Ward identity remains the same for these transformations, they
may be used to change the origin of its terms. For example, with
\be
K=\frac{1}{2}M^{-1}\{\ga,D\} \mb{,} \bar{K}=\frac{1}{2}\{\ga,D\}M^{-1} 
\label{tg}
\ee
the anomaly term of \re{gwa2} is transported from the action contribution 
to the measure contribution. 

To get the local chiral transformations one simply has to replace $\ga$ of 
the global cases by $\ga\hat{e}(n)$, where $\hat{e}(n)$ in lattice-space 
representation reads
$\big(\hat{e}(n)\big)_{n''n'}=\delta_{n''n}\delta_{nn'}$. Thus to see the
essential features it will suffice to consider the relations of the global 
case in the following.

\section{Chiral properties} 

The derivation of the identity \re{W0} implies that the occurring operators, 
acting on a vector space (with dimension number of sites times spinor dimension
times gauge-group dimension) map to this space itself. In fact, instead of the
Grassmann integrals one can equivalently consider minors and determinants, or 
generalizations thereof \cite{ke84}, for which this is a prerequisite.
The definition of adjoint operators in addition needs an inner product
so that the vector space must be a unitary one. This then allows to define 
normal operators (and their special cases as e.g.~hermitean and unitary ones) 
and is also necessary for using the notion of $\ga$-hermiticity. 

We require $D$ to be normal and $\ga$-hermitean which will be seen in the 
following to be necessary for really having chiral properties.  By normality, 
$[D,D\dg]=0$, one gets simultaneous eigenvectors of $D$ and $D\dg$. Then with 
the eigenequation 
\be
D f_k = \la_k f_k
\label{eg}
\ee
we also have 
\be
D\dg f_k = \la_k^* f_k  
\label{egdg}
\ee
where, to get the eigenvalue, also the inner product has been used. From
\re{egdg} by $\ga$-hermiticity, $D\dg = \ga D \ga$, we obtain the equation 
\be
D\ga f_k = \la_k^*\ga f_k 
\label{eg5}
\ee
which has important consequences.
The comparison of \re{eg} multiplied by $\ga$ with \re{eg5} gives
\be
[\ga,D] f_k = 0 \quad \mbox{ if } \quad \la_k \mbox{ real } 
\label{egc}
\ee
so that in the subspace of real eigenvalues of $D$ one can 
introduce simultaneous eigenvectors of $D$ and of $\ga$
\be
\ga f_k=c_k f_k  \mb{for} \la_k \mb{real}
\label{sD}
\ee
with the chirality $c_r$ taking values $+1$ and $-1$. The 
comparison of \re{eg} with \re{eg5} shows that one has simultaneously
\be
D f_k = \la_k f_k \mb{and} D \ga f_k =\la_k^* \ga f_k \mb{for} 
\la_k\ne\la_k^* \;,
\label{sC}
\ee
i.e. pairs of eigenvectors related to conjugate complex eigenvalues.

Conversely, given $\ga$-hermiticity, \re{eg5} implies normality of $D$. 
Further, having the chiral subspace, \re{eg5} follows for real eigenvalues. 
For complex ones it follows from \re{sC}. Thus to have \re{egc} and \re{sC}
normality of $D$ is also necessary.

Normality thus turns out to be necessary to get the chiral subspace which is 
the basis of chiral properties of $D$. In addition normality of $D$, being 
necessary and sufficient in order that its eigenvectors form a complete 
orthonormal set, guarantees the completeness of eigenvectors which in the 
following will be seen to be crucial for the index relations.

We note that if one would try to do without \re{sC}, one would have normality 
of $D$ only in the subspace of real eigenvalues. To specify $D$ generally in
such a way appears not feasible. Further, in the subspace of complex 
eigenvalues completeness of the eigenvectors then would only be guaranteed if 
there were no degeneracies of eigenvalues \cite{ka66}. To specify $D$ generally
such that the respective nondiagonable cases are excluded appears again not 
feasible. 
Thus we must insist that $D$ be normal in all unitary space. 

Multiplying \re{eg} from the left by $f_l\dg \ga$ and its adjoint 
$f_l\dg D\dg = f_l\dg \la_l^*$ from the right by $\ga f_k$, and using
$\ga$-hermiticity, one obtains the relation
\be
f_l\dg \ga f_k = 0 \quad \mbox{ for } \quad \la_l^* \ne \la_k \;.
\label{nela}
\ee
We note that if one makes use of the properties given by \re{sD} and \re{sC} 
the relation \re{nela} reflects the orthogonality of eigenvectors with 
different eigenvalues. This is most easily seen introducing for the 
eigenvectors $f_k$ the more detailed notations
$f_k^{(5)}$ for Im$\la_k=0$, $f_k^{(1)}$ for Im$\la_k > 0$, and
$f_k^{(2)}=\ga f_k^{(1)}$ for Im$\la_k < 0$.

With \re{sD}, \re{nela}, and the completeness of the eigenvectors of $D$ 
we obtain for the terms in the identity \re{gwa2} and for this identity itself
\be
   \lim_{\varepsilon\rightarrow 0}
\mbox{Tr}\Big((D+\varepsilon)^{-1}\ga\varepsilon \Big)= N_+(0) - N_-(0)
\label{re0}
\ee
\be
   \lim_{\varepsilon\rightarrow 0}\,
   \frac{1}{2}\mbox{Tr}\Big((D+\varepsilon)^{-1}\{\ga,D\}\Big) =
   \sum_{\la\ne 0 \mbox{ \scriptsize  real }} \Big(N_+(\la) - N_-(\la)\Big)
\label{re1}
\ee
\be
\mbox{Tr}(\ga)=
\sum_{\la \mbox{ \scriptsize real }}\Big(N_+(\la)-N_-(\la)\Big) =0
\label{res}
\ee
where the numbers of modes with chirality $\pm 1$ at a real eigenvalue 
$\la$ of $D$ are given by $N_{\pm}(\la) = 
\sum_{k\;(\la_k=\la \,\mbox{\scriptsize real})} (1\pm c_k)/2$.

It is seen that \re{re0} gives the index $N_-(0) - N_+(0)$ of $D$. The 
r.h.s.~of \re{re1} exhibits a form characteristic of the anomaly term. The sum 
rule for real modes \re{res} shows that one has the same total number of 
right-handed and of left-handed modes. The mechanism leading to a nonvanishing 
index thus is seen to work via compensating numbers of modes at different 
$\la$. 

 From \re{res} it follows that the index of $D$ can only be nonvanishing if 
a corresponding difference from nonzero eigenvalues exists. This requires 
that in addition to 0, allowing for zero modes, there must be at least one 
further real value available in the spectrum in order that the index can be 
nontrivial.  Thus it turns out that this sum rule puts severe restrictions on 
the spectrum of $D$. Obviously it is a novel manifestation of the fact 
that a nontrivial index requires breaking of the chiral symmetry.

\section{Remarks on GW relation} 

 From the general GW relation \cite{gi82} $\{\ga,D\}= D \ga R D$, using 
$\ga$-hermiticity of $D$ and $[\ga,R]=0$, one obtains
$[D,D\dg]=2D\dg[R,D]D\dg$. Therefore one should have $[R,D]=0$ in order
that $D$ gets normal which, as we have seen, is crucial for chiral 
properties and their consequences in gauge theories. Because it is necessary 
to satisfy the relation $[R,D]=0$ in a general way this means to put $R$ equal 
to a multiple of the identity. 

Thus, having to insist on normality of $D$, we 
remain with the simple form of the GW relation
\be
\{\ga,D\}= \rho^{-1} D \ga D 
\label{GW}
\ee
with $\rho$ being a real constant.
Requiring also $\ga$-hermiticity of $D$, the condition \re{GW} means that 
$\rho(D+D\dg)=D D\dg=D\dg D$ should hold, i.e.~that $D/\rho-1$ 
should be unitary. Thus the actual content of \re{GW} is the restriction of 
the spectrum of $D$ to the circle through zero with center at $\rho$. 
The crucial properties then are that real eigenvalues get 
possible at 0 and at $2\rho$, allowing for zero modes and for a nonzero 
index, respectively.

Imposing the GW relation \re{GW}, in the massless case the anomaly term in
\re{gwa2} can be expressed as 
\be
   \lim_{\varepsilon\rightarrow 0}\,
  \frac{1}{2}\mbox{Tr}\Big((D+\varepsilon)^{-1}\{\ga,D\}\Big) =
  (2\rho)^{-1}\mbox{Tr}(\ga D)
\label{rgw}
\ee
so that the identity \re{gwa2} can be replaced by
\be
\mbox{Tr}(\ga) = (2\rho)^{-1}\mbox{Tr}(\ga D) +
   \lim_{\varepsilon\rightarrow 0}
   \mbox{Tr}\Big((D+\varepsilon)^{-1}\ga\varepsilon \Big)=0 \;.
\label{gwa3}
\ee

The relation $\mbox{Tr}(\ga D)=
\sum_{\la\ne 0 \mbox{ \scriptsize real }} \la\,\Big(N_+(\la) - N_-(\la)\Big)$,
which would be not useful in the general case, now simplifies to
$\mbox{Tr}(\ga D)= 2\rho \Big(N_+(2\rho) - N_-(2\rho)\Big)$ and the sum rule 
\re{res} to $\mbox{Tr}(\ga) = N_+(0) - N_-(0)+N_+(2\rho) - N_-(2\rho)=0$.
The combination of these relations is what gives the formula 
\be
(2\rho)^{-1}\mbox{Tr}(\ga D) = N_-(0) - N_+(0)
\label{tr1}
\ee
considered in \cite{ha98,lu98} for the index. 

Using \re{GW}, one can replace \re{tg} of the alternative transformation 
transporting the anomaly term by 
$K=(2\rho)^{-1}\ga D$, $\bar{K}=(2\rho)^{-1}D\ga$. This obviously gives the 
transformation introduced in the GW case by L\"uscher \cite{lu98}, tailored to 
make the classical action $\bar{\psi} D\psi$ invariant. The measure 
contribution \re{mes} then gets $(2\rho)^{-1}\mbox{Tr}(\ga D)$. However, there 
still remains the action contribution 
$\lim_{\varepsilon\rightarrow 0}\,
\mbox{Tr}\Big((D+\varepsilon)^{-1}\ga\varepsilon \Big)$.  

The remaining action contribution is missing in \cite{lu98} since no zero-mode 
regularization has been used. Thus it looks there like the action would also
be invariant in the quantum case with zero modes, as is not correct. In a 
separate next step, which implicitly uses the decomposition \re{gwa3} of
Tr$(\ga)=0$, what should have been obtained from the action contribution 
is calculated from the measure term. This does not cure the missing in the 
originally derived identity. 

Clearly one can think of many possibilities satisfying the requirement that 
in the spectrum one should allow for at least one further real value in 
addition to 0, as imposed by the sum rule \re{res}. For finding appropriate 
constraints the decomposition 
\be
 D = u + i v \mb{with}
 u= u\dg= \frac{1}{2} (D+D\dg) \quad , \quad v= v\dg = \frac{1}{2i} (D-D\dg)
\ee
appears useful. The reason for this is that by normality of 
$D$ one obtains $\,[u,v]=0$ so that for $u$, $v$, and $D$ one gets 
simultaneous eigenvectors and the eigenvalues of $u$ and $v$ are simply
the real and imaginary parts, respectively, of those of $D$. 

For example, one may use this to define a family of constraints to a 
one-dimensional set, allowing eigenvalues at zero and at one further real 
value, by 
\be      
v^2=2\rho u + (\beta -1) u^2 \mb{with} \beta\ge 0,\; \beta\ne 1
\ee
in which case the spectrum is restricted for $\beta=0$ to the circle of the GW 
case, for $0<\beta<1$ to ellipses, and for $1<\beta$ to hyperbolas. Inserting 
$u$ and $v$ and using $\ga$-hermiticity this may be cast into the form
\be
\{\ga,D\}=\rho^{-1}\Big((1-\frac{\beta}{2})D\ga D-\frac{\beta}{4}\{\ga,D^2\}
         \Big)
\ee
which generalizes the relation \re{GW}.

\section{Continuum limit} 

For the present purpose it suffices to consider the continuum limit to the
quantum field theory of fermions in a background gauge field. The limit of 
the anomaly term in the identity \re{gwa2} has been shown to be 
\be
\frac{1}{2}\mbox{Tr}\Big((D+\varepsilon)^{-1}\{\ga,D\}\Big) \rightarrow
 -\,\frac{g^2}{32\pi^2}\int\mbox{d}^4x\;\mbox{tr}(\tilde{F}F)
\label{lim}
\ee
long ago \cite{ke81} for the Wilson-Dirac operator and recently 
\cite{ad98,su98} also for the operator of Neuberger \cite{ne98}. In the 
latter case one has to note that the l.h.s.~of \re{lim} can be replaced 
according to \re{rgw} to get the form used in \cite{ad98,su98}. 
Though there are still subtleties \cite{ad98} which deserve further
development, it can be expected that any appropriate form of $D$ should 
give \re{lim}. 

In the massless case with normal and $\ga$-hermitean $D$ we can insert
\re{re0} and \re{lim} into the identity \re{gwa2} to obtain 
\be
\mbox{Tr}(\ga)=-\,\frac{g^2}{32\pi^2}\int\mbox{d}^4x\;\mbox{tr}(\tilde{F}F)+
N_+(0) - N_-(0)=0 \;.
\label{ind}
\ee
Thus obviously the index theorem follows in the limit. To see that one also
still has $\mbox{Tr}(\ga)=0$ one has to note that any complete set of vectors 
can be used to calculate Tr$(\ga)$. In particular, one may select a set which 
exploits the fact that the spinor space factorizes off. Since in the latter 
space one has tr$(\ga)=0$, the sequence for Tr$(\ga)$ with increasing lattice 
spacing is one with all members zero, so that one has indeed $\mbox{Tr}(\ga)=0$
also in the limit. 

We emphasize that, quite remarkably, the index theorem follows here in a 
rather different setting from that of mathematics. There the Atiyah-Singer 
theorem is obtained solely considering the continuum Dirac (or Weyl) operator 
on a compact manifold finding that its index equals a topological invariant. 
Here we consider the nonperturbative formulation of the quantum field theory 
of fermions in a background gauge field and derive the chiral Ward identity. 
This identity then gives the index theorem. An essential property of this 
theory is that a chirally noninvariant modification occurs in its action. 
Additional features of the field-theoretic setting are that the Ward identity 
is a particular decomposition of Tr$(\ga)=0$ and that one gets a local version
explaining the nonconservation of the singlet axial-vector current. 

We now compare with the conventional continuum approach, in which (in our
notation) the operator $D$ is antihermitean (and thus also normal) and 
$\ga$-hermitean. Because one then has $\{\ga,D\}=0$, the anomaly term in 
the identity \re{gwa2} vanishes. However, at the level of the Ward
identity in perturbation theory (in the well known triangle diagram) one 
gets an ambiguity which, if fixed in a gauge-invariant way, produces the 
anomaly term \cite{ad69}. Thus, though with different origin of this
term, there one gets agreement with \re{ind}.

Nevertheless, there is an essential difference. While in the continuum the 
chirally noninvariant modification of the theory occurs only at the level of 
the Ward identity, on the lattice the origin of the anomaly sits in the action 
itself. Thus, since deriving things from the start is more satisfactory than 
only fixing inconsistencies later by hand, the lattice formulation is the 
preferable one. The missing of an appropriate modification at the level of 
the action in the continuum approach has the concrete consequence that there 
are difficulties with making it truely nonperturbative. 

This is seen noting that the respective attempts rely on the Pauli-Villars 
(PV) term. The motivation there is that in perturbation theory in the PV 
difference ambiguous contributions, being mass-independent, drop out so that 
the PV term gives the anomaly \cite{ad69}. Assuming the PV term to be 
nonperturbatively valid the desired result is obtained neglecting higher 
orders in the PV mass \cite{br77}.  However, one actually gets zero as 
one readily checks using Tr$(\ga)=0$, the neglect of the sum of higher orders
being not correct. This does not come as a surprise since in the lattice 
formulation it is obvious that a chirally noninvariant modification of the 
action is indispensable to get the correct result.

In the path-integral approach \cite{fu79} the usual chiral transformation 
is used so that in the global case the measure contribution is $-$Tr$(\ga)$. 
Arguing that it should be regularized this contribution, which is actually 
zero, is replaced by a term which can be checked to be equivalent to the 
PV term in \cite{br77} and from which essentially as in \cite{br77},
i.e.~incorrectly as pointed out above, the anomaly is obtained. On the other 
hand, in the Ward identity then to the mass (index) term in the action 
contribution the anomaly term is not added, as would have been necessary in 
the continuum \cite{ad69}. This compensates the unjustified replacement of 
Tr$(\ga)$ so that the desired result is obtained. 

 From our results it is obvious that to correct the procedure of 
\cite{fu79} one firstly has to use the alternative transformation with \re{tg} 
which transports the anomaly term to the measure and secondly to take care
that this term emerges properly in a nonperturbative way (requiring an 
appropriate modification of the action as is e.g.~provided by Wilson's 
regularization suppressing doublers in lattice theory). It should be 
added that the defects of the approach in \cite{fu79} also invalidate 
recent lattice considerations \cite{fu99} which rely on it.

\section{Normal $D$ from hermitean $H$} 

To get an explicit form of $D$ one can start from the Wilson-Dirac operator or 
some generalization of it, which is $\ga$-hermitean, however, not normal. The 
$\ga$-hermiticity of this operator $X$ implies that $H=\ga X$ is even hermitean
so that its spectral representation allows to define functions of $H$.
This suggests to get a normal operator $D$ from a general function of $H$ by 
imposing the necessary conditions. 

To proceed it is convenient to consider $F=\ga D$ which should generalize
$H=\ga X$. From $\ga$-hermiticity of $D$ it follows that $F$ must be hermitean 
and normality of $D$ gives the condition $[\ga,F^2]=0$. This does, however,
not yet determine $F$. In fact, with $[\ga,E^2]=0$ for some hermitean operator 
$E$ the conditions on $F$ are satisfied by $F^2=E^2+\{\ga,Y\}+c$ where $Y$ is 
some hermitean operator and $c$ some real number. From the fact that $F^2$ 
is a square it then follows that one must have $c=b^2$ nonnegative and $Y=b E$. 
One thus arrives at $F=E+b\ga$ in which the operator $E$ and the real number
$b$ are to be determined. 

With $ H \phi_l = \alpha_l \phi_l $ the definition of $E$ as a function 
of $H$ is $E(H) = \sum_l E(\alpha_l) \phi_l \phi_l\dg$ where $E(\alpha)$ is a 
real function of real $\alpha$. The task then is to determine $E(\alpha)$ in 
such a way that the condition $[\ga,E(H)^2]=0$ holds. Because $H$ does not 
commute with $\ga$ and since we are not allowed to restrict $H$ in any way 
this can only be achieved by requiring the function $E(\alpha)^2$ in 
$E(H)^2 = \sum_l E(\alpha_l)^2 \phi_l \phi_l\dg$ to be constant. Thus we get 
$E(H)^2=\rho^2\Id$ and $E(\alpha)=\pm\rho$ with $\rho$ being a real constant.

 From $E(H)^2=\rho^2\Id$ we see that $\ga E(H)/\rho$ is unitary so that the 
spectrum of $\ga E(H)$ is on a circle with radius $|\rho|$ and center at zero. 
To allow for zero modes of $D$ we therefore in $D=\ga F= \ga E+b$ have to
choose $b=\rho$ or $b=-\rho$ and cannot admit any dependence of $b$ on $H$. 
Without restricting generality taking $b=\rho$ we thus get 
$D=\rho\,(1+\ga\epsilon_0(H))$ where $\epsilon_0(\alpha)=\pm 1$. For this 
form of $D$ obviously already the GW relation \re{GW} holds. 

For $\alpha\ne 0$, requiring the function $\epsilon_0(\alpha)$ to be odd and 
nondecreasing, it gets the sign function defined by $\epsilon(\alpha)=\pm 1$ 
for $\alpha{>\atop <} 0$. That this choice is appropriate is confirmed by 
checking the classical continuum limit in the free case. Thus if all 
$\alpha_l\ne 0$ we obtain  
\be
D=\rho\,\Big(1+\ga \epsilon(H)\Big)  
\label{neu}
\ee
which is seen to be just the operator of Neuberger \cite{ne98}.

If $\alpha_l=0$ occur also $\epsilon(0)$ is to be specified. Because of the 
condition $E(H)^2=\rho^2\Id$ only either $+1$ or $-1$ is possible for this. 
To prefer none of these choices  we propose to calculate \re{neu} independently 
for each choice of $\epsilon(0)$ and to take the mean of the final results. 
As will be seen below this gives agreement with what follows from counting 
eigenvalue flows of $H$.

To the get index of $D$ by counting eigenvalue flows of $H$ has been
introduced in \cite{na93}. These flows with $m$ rely on the form 
$H(m)=H(0)+m\ga$ of the hermitean Wilson-Dirac operator and, with the 
eigenequation $ H(m) \phi_l(m) = \alpha_l(m) \phi_l(m)$, are described by the 
functions $\alpha_l(m)$. We have recently shown \cite{ke99} that these spectral
flows obey a differential equation and have given a detailed overview of the 
solutions of this equation. 

The relation to the index can be obtained inserting \re{neu} into \re{tr1} 
which in the absence of zero modes of $H$ gives 
$ N_-(0) - N_+(0)=\frac{1}{2}\mbox{Tr}(\epsilon(H))$ and in terms of numbers 
of positive and negative eigenvalues of $H$ reads 
$N_-(0)-N_+(0)=\frac{1}{2}(N_+^H-N_-^H)$. We now note that this form is also 
adequate in the presence of zero modes of $H$. In fact, following a flow, up 
to crossing there is a change by $\frac{1}{2}$ and after this a further change
by $\frac{1}{2}$. At the very moment of crossing a change of $\frac{1}{2}$ 
is reached which obviously agrees with the respective result of the procedure 
of dealing with $\epsilon(0)$ proposed above.

\section*{Acknowledgement}

I wish to thank Ting-Wai Chiu for his generous hospitality at CHIRAL '99, 
Taipei, Taiwan, Sept.~13-18, 1999, and him and all organizers for making 
this meeting such an inspiring and enjoyable one.


\end{document}